\documentstyle[epsf,twoside,fleqn,espcrc2]{article}
\newcommand{\pspicture}[1]{%
\centerline{\setlength\epsfxsize{7.25cm}\epsffile{#1}}}

\newcommand{\fig}[1]{Fig.(\ref{fig:#1})}

\newcommand{\by}{\times}
\newcommand{\sub}[1]{\mbox{\scriptsize{#1}}}
\newcommand{\ainv}{a^{-1}}
\newcommand{\GeV}{\,\mbox{GeV}}
\newcommand{\MeV}{\,\mbox{MeV}}
\newcommand{\fm}{\,\mbox{fm}}
\title{Effects of Improvement: $f_\pi$ and $m_q$}
\author{Brian Gough, Tetsuya Onogi, Jim Simone\\
\vspace{2mm}
Theoretical Physics Group, Fermilab, PO Box 500, Batavia, IL 60510, USA}
\begin{document}
\begin{abstract}
We present a study of the effects of improvement on light-quark
physics using the Fermilab formalism. The calculations were performed
at three different lattice spacings, with the SW action and
traditional Wilson action (both tadpole improved).  We find that
$O(a)$ effects for the decay constant $f_\pi$ and quark mass $m_q$ can
be successfully removed using the tadpole-improved SW action in the
light-quark regime.
\end{abstract}
\maketitle
\section{Introduction}
In this project, we have made a systematic comparison of improved and
unimproved actions by studying the lattice-spacing dependence of the
simplest quantities in light-quark physics: the pion decay constant,
$f_\pi$, and the (isospin-averaged) light-quark mass, $m_q$.

The first step in improving the Wilson action for QCD is to remove the
leading $O(a)$ errors from Wilson fermions.
Following Symanzik, this leads to the Sheikholeslami-Wohlert (SW)
$O(a)$-improved fermion action~\cite{sw:improved-action}. With
tadpole improvement~\cite{lm:viability-of-lattice-pt}, this gives a
QCD action which is essentially correct to $O(a^2)$.

A reliable improved action for light quarks would allow unquenched
calculations on small lattices, with important implications for
phenomenology.
\section{Lattice Details}
We study scaling effects using three lattices, at $\beta=5.7$,
$5.9$, and $6.1$, which cover a range of more than a factor of two in
$a$.

On each lattice we compute Wilson and SW propagators, as described
below.  In this study we aim to test the effects of Symanzik
improvement, i.e.\ the presence or absence of the SW clover
$O(a)$-improvement term in the action. We do not investigate the effects of
tadpole improvement compared with no tadpole improvement ---
tadpole improvement is used throughout, and the same
tadpole-improvement procedure is used for both actions.

The full lattice details are shown in Table~\ref{table:lattice}.
\begin{table}[t]
\caption{Lattice details ($n_f=0$)}
\label{table:lattice}
\centerline{%
\begin{tabular}{lccc}
\hline\hline
		& $\beta{=}6.1$ & $\beta{=}5.9$ & $\beta{=}5.7$	\\
$L^3\by T$	& $24^3\by 48$	& $16^3\by 32$	& $12^3\by24$	\\
$\ainv$ 	& $2.43 \GeV$ 	& $1.78 \GeV$ 	& $1.15 \GeV$	\\
$L^{phys}$ 	& $2.0 \fm$ 	& $1.8 \fm$ 	& $2.1 \fm$	\\
$c$		& $1.46^{*}$	& $1.50^{*}$	& $1.57$ 	\\
		&		&		&	\\
\# configs	&		&		&	\\
Wilson		& 100		& 100		& 200	\\
SW		& 100		& 150		& 200	\\
\# light $\kappa$'s	& 5		& 5		& 5	\\
\hline\hline	&		&		&		 
\end{tabular}}
${}^*${\footnotesize For historical reasons, the $\beta=6.1$ and $5.9$
results were obtained with the mean-field value $c=1.4$. The
differences are negligible and do not modify the conclusions.}
\end{table}
\section{Improvement Procedure}
We use a tadpole-improved SW fermion action,
\begin{equation}
S=S_{\sub{Wilson}} +(i/2)~c~{\bar \psi}\, \sigma \cdot F \, \psi
\end{equation}
where the coefficient of the clover term, $c$, is set
non-perturbatively, $c=1/(u_0)^3$.
The average link, $u_0$, is determined from the plaquette,
\begin{equation}
u_0={\left\langle(1/3)\mbox{Tr}\,U_p\right\rangle}^{1/4}
\end{equation}
to give the values of $c$ shown in Table~\ref{table:lattice}.  The
original tree-level SW action used $c=1$.

For the fermions, we take the tadpole-improved field $\tilde\psi$,
\begin{equation}
{\tilde\psi}= \sqrt{1-3\kappa/4\kappa_{crit}} \,\psi
\end{equation}
which has the correct normalization in the chiral limit, ${\tilde
\psi}=\psi/2$.

In calculating matrix elements with the SW action an improved operator
must be used~\cite{heatlie:clover-action}. In the formalism of Heatlie
{\it et al\,} this is equivalent to rotating the quark fields,
\begin{equation}
\psi \to \left[ 1- (z \not\!\! D-(1-z)m_0)/2 \right] \,\psi.
\end{equation}
Choosing $z=0$, the rotation becomes,
\begin{equation}
\psi \to \left[ 1+m_0/2 \right] \,\psi
\end{equation}
and its effects will vanish in the chiral limit, where $m_0=0$.

With the Fermilab formalism the corresponding rotation is,
\begin{equation}
\psi \to \sqrt{1+m_0} \left[ 1 + d_1 {\vec D} \cdot {\vec \gamma} 
\right] \,\psi
\end{equation}
but its effects also vanish in the chiral limit (since $d_1\sim m_0$
for small $m_0$). For convenience we used this trick to obtain our
decay-constant results from the ordinary unrotated fields.

In chiral extrapolations we use linear fits in the lattice pole mass,
\begin{equation}
m_q=\ln\left(1+\left( 1/2\kappa - 1/2\kappa_{crit} \right) \right)
\end{equation}
to model the $\kappa$-dependence. This definition automatically resums
the tree-level large-$m$ corrections in the quark mass.

For the Wilson action we use exactly the same tadpole-improvement
procedure for the fermions, but set $c=0$.
\section{Setting the Scale}
In a systematic study of $O(a)$ effects it is important to choose a
scale which is as free from $a$-dependence as possible, so that other
quantities are not affected by a common variation in the scale.

We set the scale using the spin-averaged $1P-1S$ splitting from
charmonium~\cite{el-khadra:alphas}. The spin-averaged splitting is
found to be insensitive to $O(a)$ corrections from the clover term,
and the resulting scale is free from $O(a)$ effects. Using this scale,
the $a$-dependence of other quantities can be determined directly, up
to the remaining $O(a^2)$ systematic errors in the scale.

The advantage of using charmonium to set the scale is that it is
possible to separate the $a$-dependence of the quantities under study
from the $a$-dependence of the scale (known to be good to
$O(a^2)$). Other possible scales, such as $m_\rho$, have the
disadvantage that the scale itself can contain an $O(a)$ effect.
\section{Multistate Fitting}
When studying small $O(a)$ effects it is important to keep systematic
errors under control. These errors may have an $a$-dependence which
conceals the true scaling behavior of the quantities under study. In
particular, excited-state contamination can introduce large
$O(a)$ effects, purely from the systematic variations in a poor
fitting procedure.

We use a sophisticated multistate fitting program to eliminate
excited-state contamination. A set of optimised smearing functions is
available on each lattice, created from a study of coulomb-gauge pion
wavefunctions, typically for $1S$ and $2S$ sources. The fitting is
performed using a matrix of correlators, $\delta$-$\delta$ (point
source), $1S$-$1S$, $2S$-$2S$, $A_4$-$\delta$,
\dots, and a choice of one-exponential or two-exponential fits. We
check for consistency between the different fitting methods, to ensure
that excited-state contamination has been removed.
\section{Decay Constant}
\begin{figure}[t]
\pspicture{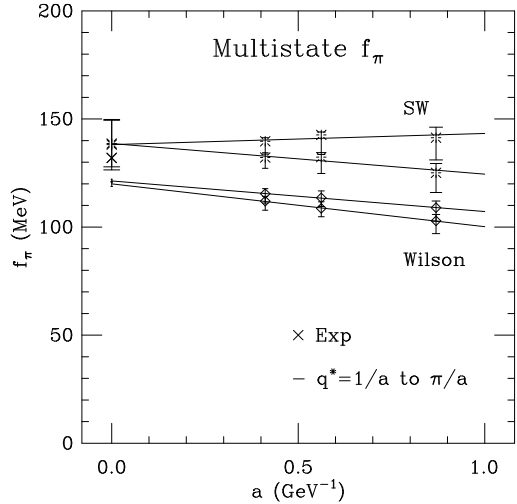}
\caption{Comparison of SW and Wilson $f_\pi$. 
The lines show the effect of varying $q^*$ from $1/a$ (lower line) to
$\pi/a$ (upper line).}
\label{fig:fpi}
\end{figure}
Previous studies of $f_\pi$ using Wilson fermions have all shown a
negligible dependence on lattice-spacing. In order to determine
whether this is due to an intrinsically small $O(a)$ correction for
$f_\pi$ from the action, or a cancellation by $O(a)$ effects from
excited-state contamination, our analysis was carried out using
multistate fits.

Fermilab multistate results for $f_\pi$ using the tadpole-improved
Wilson and tadpole-improved SW actions are shown in \fig{fpi}, with
the physical value $f_\pi=132\MeV$. The lattice values are also
extrapolated to the continuum limit using a simple linear fit in $a$.

The Wilson results are renormalized using the tadpole-improved $Z_A$
of Lepage and Mackenzie~\cite{lm:viability-of-lattice-pt}. The
tadpole-improved value of $Z_A$ for the local current with the SW
action was obtained from the work of Borrelli {\it et
al}\,~\cite{borrelli:improved-operators}, by subtracting the tadpole
contribution from their intermediate results,
\begin{eqnarray}
Z_A(q^*) &=& 1-0.31\alpha_V(q^*)~\mbox{(Wilson)}\\ 
	 &=& 1-0.61\alpha_V(q^*)~\mbox{(SW)}.
\end{eqnarray}
The exact values of $q^*$ remain to be determined by analytic
calculations. For the moment, we allow $q^*$ to vary over a reasonable
range, from $1/a$ to $\pi/a$, and regard this as a systematic error.

The additional systematic errors from the $1P-1S$ charmonium scale are
expected to be $O(p^2 a^2)$, which at $\beta=5.7$ might be $5$-$10\%$.

Allowing for these systematic errors, the multistate fits confirm the
behavior of the existing Wilson data, and show a small lattice-spacing
dependence. This indicates that the coefficient of the $O(a)$
correction to Wilson $f_\pi$ is intrinsically small.

The SW results are consistent with a small or no $O(a)$ effect within
systematic errors. The addition of the $O(a)$ correction term to the
Wilson action does not increase $a$-dependence of the results when the
$a$-dependence is already small. This is a requirement of improvement.

Even allowing for the uncertainty in $q^*$, the linear extrapolations
of Wilson and SW $f_\pi$ to the continuum limit do not meet. This is a
possible indication of the perturbative ($g^4$) or quadratic ($a^2$)
corrections to the naive linear scaling law.
\section{Quark Mass}
\begin{figure}[t]
\pspicture{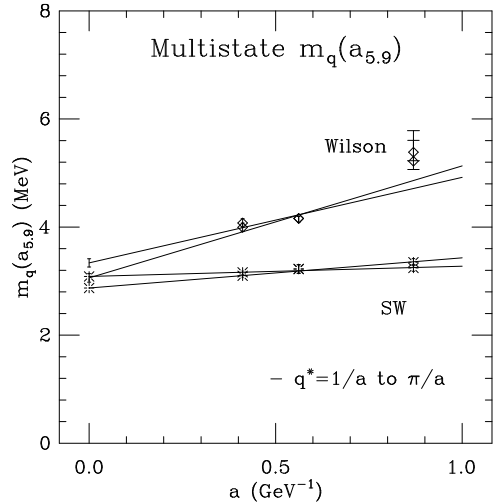}
\caption{Comparison of SW and Wilson $m_q$.
The lines show the effect of varying $q^*$ from $1/a$ 
(lower line at large $a$) to $\pi/a$ (upper line at large~$a$).}
\label{fig:mq}
\end{figure}
We extract the light-quark mass from the pion mass chiral
extrapolation (by determining the quark mass that gives the physical
pion mass),
\begin{equation}
m^2_\pi = A m_q .
\end{equation}
In Gupta's Lattice '94 review of quark masses a compilation of
existing Wilson data showed a large variation of $m_q$ with
lattice-spacing, and a significant discrepancy between Wilson and
Staggered results~\cite{gupta:chiral-limit}.

Again, in order to determine whether this is due to systematic
$O(a)$ effects from excited-state contamination, our analysis was
carried out using multistate fits.

The Fermilab multistate results for $m_q$ using tadpole-improved
Wilson and tadpole-improved SW actions are shown in \fig{mq}.  The
lattice $m_q$ values are converted to a common scale of $\beta=5.9$
using the 1-loop relation,
\begin{equation}
m(a_{5.9})=\left(1-(2/ \pi) \alpha(q^*) \ln(a/a_{5.9}) \right)
		 m(a).
\end{equation}
This removes the lattice-spacing dependence due to the running of the
mass. We keep the masses in the lattice scheme. An additional constant
factor, $Z_{\overline{MS}}$, would be needed to obtain the
conventional $\overline{MS}$ value. Since we are only interested in
the scaling properties this overall constant is not important here.

The masses are again extrapolated to the continuum limit using a
simple linear fit in $a$. In this case the Wilson and SW
extrapolations are in good agreement.

Multistate fitting confirms the presence of a large $O(a)$ effect in
Wilson $m_q$. The SW results show a small $O(a)$ effect, compatible
with zero within systematic errors. This indicates that the
discrepancy between the existing Wilson and Staggered results is
probably due to a large $O(a)$ error from the Wilson action. The
Staggered action, which is also correct to $O(a^2)$, should be in
agreement with the tadpole-improved SW action.
\section{Conclusions}
For the decay constant, multistate fitting confirms the small $O(a)$
effects found in Wilson $f_\pi$. The $O(a)$ effects for the improved
SW $f_\pi$ are also small, within systematic errors from $q^*$ and
$p^2 a^2$ corrections to the charmonium scale. There is a difference
between the continuum extrapolations of SW and Wilson $f_\pi$ which
suggests perturbative or quadratic corrections to naive linear
scaling.

For the quark mass, multistate fitting confirms the large
$O(a)$ effects found in Wilson $m_q$. The $O(a)$ effects for the SW
$m_q$ are consistent with zero, within systematic errors. The improved
action removes the $a$-dependence of $m_q$ even down to $\beta=5.7$,
when using the clover term with tadpole improvement.

Further simulations at lower $\beta$ values are desirable, to
determine the scale at which $O(a^2)$ errors become important. At this
stage there is no clear evidence of $O(a^2)$ effects, even at
$\beta=5.7$. However, for larger lattice spacings it is essential to
first obtain the correct $q^*$ values, since the systematic error due
to $q^*$ increases as the perturbative corrections increase.

The costs involved in using the SW action are small compared with the
benefits of reducing $O(a)$ errors. On the {\sc canopy/acpmaps}
platform the addition of the clover term typically increases the
runtime of the quark inversion by approximately 30\%.

In summary, when Wilson results already show small $O(a)$ effects,
improvement does not make things worse, and when Wilson results show
large $O(a)$ effects, improvement reduces them. Thus, use of the
tadpole-improved SW action is recommended.
\section{Acknowledgments}
We wish to thank our collaborators, and our colleagues in the Fermilab
Computing Division.  These calculations were performed on the Fermilab
{\sc acpmaps} supercomputer.  Fermilab is operated by Universities Research
Association, Inc., under contract DE-AC02-76CH03000 with the
U.S. Department of Energy.
\newcommand{\noopsort}[1]{} \newcommand{\printfirst}[2]{#1}
  \newcommand{\singleletter}[1]{#1} \newcommand{\switchargs}[2]{#2#1}

\end{document}